\documentclass[dvips]{article}
\usepackage{amssymb,amsmath,latexsym, mathrsfs}
\usepackage[dvips]{color}
\usepackage{graphicx} 

\renewcommand{\thesection}{\arabic{section}}

\renewcommand{\appendix}[1]{\setcounter{section}{0}
 \setcounter{equation}{0}
 \renewcommand{\theequation}{{#1}.\arabic{equation}}
 \renewcommand{\thesection}{}}

\newcommand{\beq}[1]{  \\ {\tiny ({#1})}   \begin{equation} \label{#1} }
\newcommand{\bal}[1]{ \\ {\tiny ({#1})}   \begin{align} \label{#1} }
\renewcommand{\beq}[1]{  \begin{equation} \label{#1} }    
\renewcommand{\bal}[1]{\begin{align} \label{#1} }		
\newcommand{\eeq}{\end{equation}}

\newcommand{\rf}[1]{(\ref{#1})}

\def\grad{\operatorname{grad}} 
\def\div{\operatorname{div}} 

\def\bd#1{\ensuremath{\mathchoice
                     {\mbox{\boldmath$\displaystyle\mathbf{#1}$}}
                     {\mbox{\boldmath$\textstyle\mathbf{#1}$}}
                     {\mbox{\boldmath$\scriptstyle\mathbf{#1}$}}
                     {\mbox{\boldmath$\scriptscriptstyle\mathbf{#1}$}}}}

\def\doublespacing{\baselineskip=18pt}
\begin{document}
\doublespacing
\begin{center}

{\huge Impedance of a sphere oscillating  in an elastic medium with and without slip }

\bigskip

Andrew N. Norris

\medskip
Department of Mechanical \& Aerospace
Engineering, \\Rutgers University, 
98 Brett Road, Piscataway, NJ
08854-8058, USA

\end{center}

\begin{center} {\Large Abstract}\end{center}

The dynamic impedance of a sphere oscillating in an elastic medium is considered.  Oestreicher's \cite{Oestreicher51} 
formula for the impedance of a sphere bonded to the surrounding medium can be expressed simply in terms of three lumped impedances associated with the displaced mass and the longitudinal and transverse waves.  If the surface of the sphere slips while the normal velocity remains continuous, the impedance formula is modified by adjusting the definition of the transverse impedance to include the interfacial impedance. 

\medskip \noindent

\vskip 0.5in
\noindent
Short title: Thermoelastic thin plates\\

\medskip\noindent
Contact: \\
norris@rutgers.edu\\ 
ph. 1.732.445.3818\\
fax 1.732.445.3124

\newpage 
\section{Introduction}


The dynamic impedance of a spherical particle embedded in a medium is important for acoustical measurement and imaging.  The impedance is used, for instance,  in  measurement of the mechanical properties of tissue, e.g. \cite{Yin04}, and  is  intimately related to the radiation forcing on particles \cite{Chen02}.  The latter is the basis for imaging techniques such as vibro-acoustography 
which has considerable potential in mammography for detection of  microcalcifications in breast tissue
\cite{Fatemi02}. 
Oestreicher  \cite{Oestreicher51}  derived the  impedance for a rigid sphere oscillating in a viscoelastic medium over 50 years ago.  Although it was derived for a sphere in an infinite medium, Oestreicher's formula is also  applicable, with minor modification \cite{Gierke52}, to dynamical indentation  techniques where the particle is in contact with the surface of the specimen, see Zhang et al. \cite{Zhang01} for a review of related work.  
Chen et al. \cite{Chen05} recently   validated Oestreicher's formula experimentally by measuring the  dynamic radiation force on a sphere in a fluid.   The impedance formula is based on perfect no slip conditions between the spherical inclusion and its surroundings.  This may not always be a valid assumption,  e.g., in circumstances where a foreign object is embedded in soft material.  This was precisely the situation in recent measurements of the viscosity of DNA cross-linked gels by magnetic forcing on a small steel sphere  \cite{Lin04}.  This paper generalizes the impedance formula to include the possibility of dynamic slip. 

Two related results are derived in this paper.  The first is a modified form of Oestreicher's formula which enables it to be interpreted in terms of lumped parameter impedances.  This leads to a simple means to consider the more general case of a sphere oscillating in a viscoelastic medium which is permitted to slip relative to its  surroundings.  The slip is characterized by an interfacial impedance which relates the shear stress to the discontinuity in tangential velocities. This generalization includes Oestreicher's original formula as the limit of infinite interfacial impedance, and agrees with previous results for the static stiffness of a spherical inclusion with and without slip \cite{Lin05}.

\section{Summary of results}

A  sphere  undergoes time harmonic oscillatory motion  of  amplitude $u_0$ in the direction  $\hat{\bd x}$, 
\beq{01}
{\bd u}^{\text sphere} = u_0 \, e^{-i \omega t}\, \hat{\bd x}. 
\eeq
The time harmonic factor $e^{-i \omega t}$ is  omitted but understood in future expressions.
The sphere, which is assumed to be rigid  and of radius $a$, is embedded in an elastic medium of infinite extent with  mass density $\rho$ and Lam\'e moduli $\lambda$ and $\mu$. The moduli may be real or complex, corresponding to an elastic or viscoelastic solid.  We will later consider complex 
 shear modulus  $\mu = \mu_1 -i\omega \mu_2$, where the imaginary term dominates in a viscous medium.  The force exerted on the sphere by the surrounding medium acts in the $\hat{\bd x}$-direction, and is defined by 
\beq{04}
F\hat{\bd x}=  \int\limits_{r=a} {\bd T}\, {\text d} s\,  , 
\eeq
where ${\bd T}$ is  the  traction vector on the surface. 
The sphere impedance is defined
\beq{11}
Z = \frac{F}{-i\omega u_0}. 
\eeq
 Oestreicher's expression for the impedance of a sphere that does not slip relative to the elastic medium is\footnote{Equation \rf{13.5} is 
 Oestreicher's \cite{Oestreicher51} eq. (18) with $i$ replaced by $-i$ since he used time dependence $e^{i\omega t}$.}  \cite{Oestreicher51} 
\begin{align}\label{13.5}
& Z = \frac43 \pi a^3\rho \,i\omega \,   \bigg[ 
\bigr( 1+\frac{3i}{ah}- \frac{3}{a^2h^2}\bigr) - 
2\bigr( \frac{-i}{ah}+ \frac{1}{a^2h^2}\bigr)\bigr( 3- \frac{a^2k^2}{1-i ak} \bigr)
\bigg] \bigg/ 
\nonumber 
\\ & \qquad\qquad\qquad\qquad \bigg[
\bigr( \frac{-i}{ah}+ \frac{1}{a^2h^2}\bigr) \frac{a^2k^2}{1-i ak} +
\bigr( 2- \frac{a^2k^2}{1-i ak} \bigr)\bigg] \, . 
\end{align}
Here $k$ and $h$ are, respectively, the longitudinal and transverse wavenumbers, $k=\omega/c_L$, $h=\omega/c_T$ with $c_L = \sqrt{(\lambda + 2\mu)/\rho }$ and $c_T = \sqrt{\mu/\rho }$. 

Noting  that Oestreicher's formula can be rewritten
\beq{13}
Z = \frac43 \pi a^3\rho \,i\omega \,   \bigg\{ -1 + \bigg[ 
\frac13\bigr( 1 - \frac{3(1-ika)}{k^2a^2}\bigr)^{-1}
+\frac23\bigr( 1 - \frac{3(1-iha)}{h^2a^2}\bigr)^{-1}
\bigg]^{-1} \bigg\}\, ,  
\eeq
implies our first result,  that  the impedance satisfies 
\beq{17}
\frac3{Z + Z_{\text m}} = \frac1{Z_L + Z_{\text m}} +\frac2{Z_T + Z_{\text m}}\, , 
\eeq
where the three additional impedances are defined as
\begin{subequations}\label{16}
\begin{align}
Z_{\text m} &=    i\omega\, \frac43 \pi a^3\rho  ,  
\\
Z_L &=   ( i\omega)^{-1} 4 \pi a(\lambda+2\mu) \, (1-ika),
\label{16b}
\\
Z_T &=   ( i\omega)^{-1}  \, 4 \pi a\mu (1-iha).
\label{16c}
\end{align}
\end{subequations}

The second result is that if the  sphere is allowed to  slip relative to the elastic medium then  the general form of eq. \rf{17} is preserved with $Z_T$ modified.  Specifically, suppose the tangential component of the traction satisfies 
\beq{67}
{\bd T}\cdot \hat{\bd t} = z_I\, ( \dot{\bd u}^{\text sphere}- {\bd v})\cdot \hat{\bd t} ,
\qquad r=a,
\eeq
where $ \hat{\bd t}$ is a  unit tangent vector, $ {\bd v}$ the velocity of the elastic medium adjacent to  the sphere, and $z_I$ is an interfacial impedance\footnote{Capital $Z$ and lower case $z$ are used to distinguish impedances defined by force and stress, respectively.}, discussed later.  Equation \rf{67} holds at each point on the interface $r=a$.  We find that $Z$ now satisfies
\beq{18.1}
\frac3{Z + Z_{\text m}} = \frac1{Z_L + Z_{\text m}} +\frac2{Z_S + Z_{\text m}}\, , 
\eeq
where the new impedance $Z_S$ is   given by 
\beq{18.2}
\frac1{Z_S}= \frac1{Z_T} + \frac1{  4 \pi a^2 z_I + (i\omega)^{-1}  \, 8 \pi a\mu }\, . 
\eeq
These results are derived in the next Section and discussed in Section \ref{Discussion}. 

\section{Analysis}

We use  Oestreicher's  \cite{Oestreicher51} representation for the  elastic field outside the sphere, 
\beq{02}
{\bd u} = -A_1 \grad \big( \frac{h_1(kr)}{kr}x\big) 
+ B_1\big[ 2h_0(hr)\grad x - h_2(hr)r^3\grad \frac{x}{r^3}\big], \qquad r\ge a, 
\eeq
where $r = |{\bd r}|$ is the spherical radius  and  $x$ is the component of ${\bd r}$ in the $\hat{\bd x}-$direction,  both with origin at the center of the sphere. Also,
$h_n$ are spherical Hankel functions of the first kind \cite{Abramowitz74}.   Let $\hat{\bd r} = r^{-1}{\bd r}$ denote the unit radial vector, then 
\beq{03}
{\bd u} = -A_1  \big[ \frac{h_1(kr)}{kr}\hat{\bd x} -h_2(kr) \frac{x}{r}\hat{\bd r}
\big]
+ B_1\big[ 2h_0(hr)\hat{\bd x} - h_2(hr) (\hat{\bd x} - 3 \frac{x}{r}\hat{\bd r} )\big]. 
\eeq
The  surface traction is ${\bd T} = {\bd \sigma}\hat{\bd r} $ where ${\bd \sigma}$ is the stress tensor. 
 The traction  can be calculated from  \rf{03} and the following identity \cite{Oestreicher51} for an isotropic solid,
\beq{05}
  {\bd T}  = \hat{\bd r} \lambda \div  {\bd u} + \frac{\mu}{r} \grad {\bd r}\cdot {\bd u}
 + \mu \big(  \frac{\partial }{\partial r} - \frac1{r}\big) {\bd u}\, . 
\eeq
Thus, referring to \rf{04}, we have 
\bal{051}
  {\bd T} \cdot \hat{\bd x}  = &
  \big[  2\mu h_2(kr) \big(1-3 \frac{x^2}{r^2}\big)   +
  (\lambda + 2\mu ) kr h_1(kr) \frac{x^2}{r^2}  \big]\, \frac{A_1}{r}
  \nonumber \\
  & + \big[ 2 h_2(hr) \big(1-3 \frac{x^2}{r^2}\big)   -
   hr h_1(hr) \big(1- \frac{x^2}{r^2}\big) \big]\, 3\mu \frac{B_1}{r}\, . 
\end{align}
Integrating over the sphere surface, the resultant is 
\beq{06}
F = \frac43 \pi a^3 \rho \omega^2\, \big[ 
A_1 \frac{h_1(ka)}{ka} - 6B_1 \frac{h_1(ha)}{ha} \big]\, . 
\eeq

The coefficients $A_1$ and $B_1$ follow from the conditions describing the interaction of the sphere with its surroundings.  These are the  general slip condition \rf{67} plus the requirement that the normal velocity is continuous.  The conditions at the surface of the sphere are 
\beq{091}
\begin{split}
{\bd u} \cdot\hat{\bd r}  &= u_0 \, \hat{\bd x}\cdot  \hat{\bd r}  
\\
{\bd T} \cdot\hat{\bd t}  &= i\omega z_I\, ( {\bd u} -u_0 \hat{\bd x})\cdot \hat{\bd t}
\end{split} 
\, \, 
\Biggr\} 
\qquad r= a . 
\eeq
By symmetry,  the only non-zero  tangential component is in the plane of $\hat{\bd r}$ and $\hat{\bd x}$, and we therefore set $\hat{\bd t}= \hat{\bd \theta} \equiv (\hat{\bd r} \cos\theta - \hat{\bd x})/\sin\theta $ where $ \theta =\arccos \hat{\bd r}\cdot \hat{\bd x}$ is the spherical polar angle.  Using polar coordinates, ${\bd u} = u_r \hat{\bd r} + u_\theta \hat{\bd \theta}$ and  ${\bd T} =\sigma_{rr}\hat{\bd r} +\sigma_{r\theta}\hat{\bd \theta}$, and \rf{091} becomes
\beq{092}
\begin{split}
u_r &= u_0 \,\cos\theta 
\\
\sigma_{r\theta} - i\omega z_I u_\theta &= i\omega z_I u_o \sin\theta 
\end{split} 
\, \, 
\Biggr\} 
\qquad r= a,\quad 0\le \theta \le \pi . 
\eeq
The shear stress follows from  the identity 
\beq{093}
\sigma_{r\theta} = \mu\big( \frac{\partial u_\theta}{\partial r}
+ \frac1r \frac{\partial u_r}{\partial \theta} - \frac{u_\theta}{ r}\big) , 
\eeq
 and the interface conditions \rf{092} then imply, respectively,
\begin{subequations}\label{101}
\begin{align}
 \big[ h_2(ka) -  \frac{h_1(ka)}{ka}\big] \, A_1 + 6\frac{h_1(ha)}{ha}\, B_1  &=u_0. 
\\
- \big[  \frac{h_1(ka)}{ka} + \frac{2\mu h_2(ka)}{i\omega a z_I}\big] \, A_1 
+  \big[ (2 + \frac{\mu h^2a^2}{i\omega a z_I} ) \frac{h_1(ha)}{ha} - (1+\frac{2\mu}{i\omega a z_I} )h_2(ha)
 \big] \, 3B_1 &=u_0.
\end{align}
\end{subequations}
Solving for $A_1$ and $B_1$, then substituting them into eqs. \rf{06} and \rf{11}, and using the known forms for the spherical Hankel functions,  yields 
\beq{18.3}
Z = -  Z_{\text m}+ 3/\big[ 1/(Z_L + Z_{\text m}) +2/(Z_S + Z_{\text m})\big]\, . 
\eeq
Equation \rf{18.3} is identical to \rf{18.1}, which completes the derivation of the generalized impedance formula. 

\section{Discussion}			\label{Discussion}

It is useful to recall some basic properties of lumped parameter impedances. 
The impedance of a  spring mass damper system of stiffness $K$, mass $M$ and damping $C$ is
\beq{23}
Z = ( i\omega)^{-1} K- C +  i\omega M\, . 
\eeq
Two impedances $Z_1$ and $Z_2$  combined in series have  an effective impedance  $(Z_1^{-1}+Z_2^{-1})^{-1}$, while the result for the same pair   in parallel is  $(Z_1+Z_2)$. 

Referring to the definitions of eq. \rf{16}, it is clear that $Z_{\text m}$ is the  impedance of the mass of the  volume removed from the elastic medium.   The impedance of a longitudinal or transverse plane wave is defined as the ratio of the stress (normal or shear) to particle velocity, and equals $z_{L}$, $z_{T}$, where 
\beq{312}
z_{L}= \rho c_{L}, \qquad z_{T}= \rho c_{T}.  
\eeq
Thus, both $Z_L$ and $Z_T$ have the form 
\beq{313}
Z= \big( \frac1{i\kappa a} - 1\big) \, 4\pi a^2\, z ,  
\eeq
where $\kappa$ is the wavenumber ($k$ or $h$). In particular, the impedances $Z_L$ and $Z_T$  have stiffness and damping, but no mass contribution.  The damping can be ascribed to the radiation of longitudinal and transverse waves from the sphere. 

The impedance $Z_S$ of eq. \rf{18.2} corresponds to $Z_T$ in series  with an impedance $Z_I$, where 
\beq{555}
Z_I =  4 \pi a^2 z_I + \frac{ 8 \pi a^2 z_T}{iha}\, . 
\eeq
Thus, 
$Z_I$ can be interpreted as the total interfacial impedance for the surface area of the sphere 
in parallel with twice the stiffness part of $Z_T$.

The limit of a purely acoustic fluid is obtained by letting the shear modulus $\mu$ tend to zero with $\lambda$ finite, 
while an incompressible elastic or viscous medium is obtained in the limit as the bulk modulus $\lambda + \tfrac23 \mu$  becomes infinite with $\mu$ finite.  The acoustic and incompressible limits follow from \rf{18.3} as 
\beq{18}
Z = 
\begin{cases}
\big( \frac2{ Z_{\text m}} + \frac3{ Z_L}\big)^{-1}, & \qquad  \text{acoustic medium}, \\
& \\
\tfrac12 { Z_{\text m}} + \tfrac32 { Z_S} , & \qquad  \text{incompressible medium}.
\end{cases}
\eeq
Thus, $Z$ for the acoustic fluid comprises $\tfrac12 Z_{\text m}$ in series with  $\tfrac13 Z_L$.  Note that, as expected, the interfacial impedance $z_I$ is redundant in the acoustic limit.   The impedance for the  incompressible medium is  $\tfrac12 Z_{\text m}$ and 
$\tfrac32 Z_S$ in parallel, and it depends upon the interfacial impedance.  

In order to examine the role of $z_I$, we first express
the impedance $Z$ of eq. \rf{18.3}  in a form similar to \rf{13}, 
\beq{131}
Z = \frac43 \pi a^3\rho \,i\omega \,   \bigg\{ -1 + \bigg[ 
\frac13\bigr( 1 - \frac{3(1-ika)}{k^2a^2}\bigr)^{-1}
+\frac23\bigr( 1 - \frac{3(1-iha)}{h^2a^2[1+\frac{\chi}{2} (1-iha)]}\bigr)^{-1}
\bigg]^{-1} \bigg\}\, ,  
\eeq
where the influence of the interfacial impedance is  represented through the  non-dimensional parameter 
\beq{738}
\chi = \big( 1 + \frac{i\omega a z_I}{2\mu}\big)^{-1}  \, .  
\eeq
The form of  $\chi$  is chosen so that it takes on the values zero or unity in the limit that the sphere is perfectly bonded or is perfectly lubricated, 
\beq{12}
\chi = 
\begin{cases}
0, & \text{no slip}, \quad z_I\rightarrow \infty, \\
1, & \text{slip}, \quad \quad \  z_I = 0.
\end{cases}
\eeq
The acoustic and incompressible limits of \rf{18} are explicitly
\beq{185}
Z = 
\begin{cases}
\frac43 \pi a^3\rho \, i\omega\, \frac{(1-ika)}{2(1-ika)- k^2a^2} , & \qquad  \text{acoustic}, \\
& \\
\frac{6 \pi a \mu }{ i\omega}\, \big[ \frac{ 1-iha}{1+\frac{\chi}{2} (1-iha)} - \frac{h^2a^2}{9}\big] , & \qquad  \text{incompressible}.
\end{cases}
\eeq
Oestreicher  \cite{Oestreicher51}  showed that the original formula \rf{13.5} provides the acoustic and incompressible limits for perfect bonding $(\chi = 0)$. Ilinskii et al. \cite{Ilinskii05}  derived the impedance in the context of incompressible elasticity, also for the case of no slip.  

The  behavior of $Z$ at low and high frequencies depends upon how $z_I$ and hence $\chi$ behaves in these limits.  For simplicity, let us consider $\chi$ as constant in each limit, equal to  $\chi_0$ at low frequency, and  $\chi_\infty$ at high frequency. Then, 
\beq{19}
Z = 
\begin{cases}
( i\omega)^{-1}  \, \frac{12\pi a \mu  }{2+ \chi_0 +{c_T^2}/{c_L^2}}\, 
\big[ 1 - i ha \, \big( \frac{2+ {c_T^3}/{c_L^3} }{2+ \chi_0 +{c_T^2}/{c_L^2}}
\big)  + {\text O}(h^2a^2)\big]\, , \quad |ha|, |ka|\ll 1, 
\\  \\ 
 \frac43 \pi a^2 \rho c_L 
\bigr[ - \big( 1+2  \tfrac{c_T}{c_L} (1-\chi_\infty) \big) 
 \, +\frac{1}{ika}   \big( 1- 4 \tfrac{c_T}{c_L} [ 1+  ( \tfrac{c_T}{c_L} -1)\chi_\infty ] \big) 
 \\ 
 \qquad  \qquad  \qquad  \qquad  \qquad  \qquad  \qquad  \qquad \quad + {\text O}\big( \frac1{k^2a^2} \big) \bigr]\, ,\quad  |ha|, |ka|\gg 1.
\end{cases}
\eeq
The leading order term at high frequency is a damping, associated with radiation from the sphere.  The dominant effect at low frequency is, as one might expect, a stiffness, with the second term a damping.  The low frequency stiffness is identical to that previously determined by Lin et al 
\cite{Lin05} who considered the static problem of a sphere in an elastic medium with an applied force.  They derived the resulting displacement, and hence stiffness, under slip and no slip conditions. 
In order to compare with their results, we rewrite the leading order term as
\beq{19a}
Z = ( i\omega)^{-1}  \, \frac{24\pi a \mu (1-\nu) }{5-6\nu + 2(1-\nu) \chi_0 } \big[ 1 + {\text O}\big( 1\big) \big] , 
\eeq
where $\nu$ is the Poisson's ratio, 
\beq{20a}
\nu = \frac{\frac12 c_L^2 - c_T^2}{c_L^2 - c_T^2} .  
\eeq
Equation \rf{19a} with $\chi_0=0$ and $\chi_0=1$ agrees with eqs. (40) and (41) of Lin et al. \cite{Lin05}, respectively.  In an incompressible  viscous medium  with $\nu \approx 1/2$ and $\mu = -i\omega \mu_2$, \rf{19a} becomes
\beq{19b}
Z \approx  - \frac{6\pi a \mu_2  }{1+\tfrac12 \chi_0 } , 
\eeq
which reduces to the Stokes \cite{Stokes51} drag formula $F= -6\pi a \mu_2v$  for perfect bonding.  When there is slip ($\chi_0=1$) the drag is reduced by one third, $F= -4\pi a \mu_2v$.  It is interesting to note that one third of the  contribution to the drag in Stokes' formula is from pressure, $2\pi a \mu_2v$, the remained from shear acting on the sphere.  However, under slip conditions, the shear force is absent and the total drag $4\pi a \mu_2v$ is caused by the pressure. 

The simplest example of the interfacial impedance  is a constant value, which is necessarily negative and corresponds to a damping, $z_I = -C$. For an elastic medium we have 
\beq{445}
\chi = \frac1{1-i\omega/\omega_c}, \qquad 
\omega_c  = \frac{2\mu}{a C}, \qquad \text{ elastic medium}, \, \, z_I = -C. 
\eeq
Hence,  $\chi_0 = 1$ and $\chi_\infty = 0$, corresponding to slip at low frequency and no slip at high frequency.   The transition from the low to high frequency regime  occurs for frequencies in the range of  a characteristic frequency $\omega_c$. 
Alternatively, if the medium is purely  viscous  $\mu = -i\omega \mu_2$, 
 again with constant $z_I$, the parameter $\chi$ becomes 
\beq{446}
\chi = \frac1{1+\frac{a C}{2\mu_2}}, \qquad \text{ viscous medium}, \, \,\mu =  -i\omega \mu_2, \, \,  z_I = -C. 
\eeq
In this case $\chi$ is constant with a value between $0$ and $1$ that depends upon the ratio of the interfacial to bulk viscous damping coefficients, and also upon $a$.  One can define a characteristic particle size $a_c = \mu_2/C$, such that spheres of radius $a\ll a_C$ ($a\gg a_C$) are effectively bonded (lubricated).   

Figures 1 and 2 show the reactance and resistance of a sphere of radius $0.01$ m in a medium with the   parameters considered by Oestreicher \cite{Oestreicher51} based on measurements of human tissue, $\rho = 1100\,$ kg/m$^3$, $\mu_1=2.5\times10^3 \, $Pa, $\mu_2=15 \, $Pa{\ }sec, $\lambda=2.6\times10^9 \,$ Pa.  The perfectly  bonded ($\chi = 0$) and perfect slip ($\chi = 1$) conditions are compared.  Figure 1 indicates that the mass-like reactance is generally reduced by the slipping, and it also shows that the low frequency stiffness is two-thirds that of the bonded case, eq. \rf{19a}.  Interfacial slip leads to a significant  decrease in the resistance,  as evident from Figure 2  which shows a reduction for all frequencies.  

\newpage

\newpage
\begin{figure}[htbp]
		\begin{center}	
				\includegraphics[width=5in, height=4 in 					]{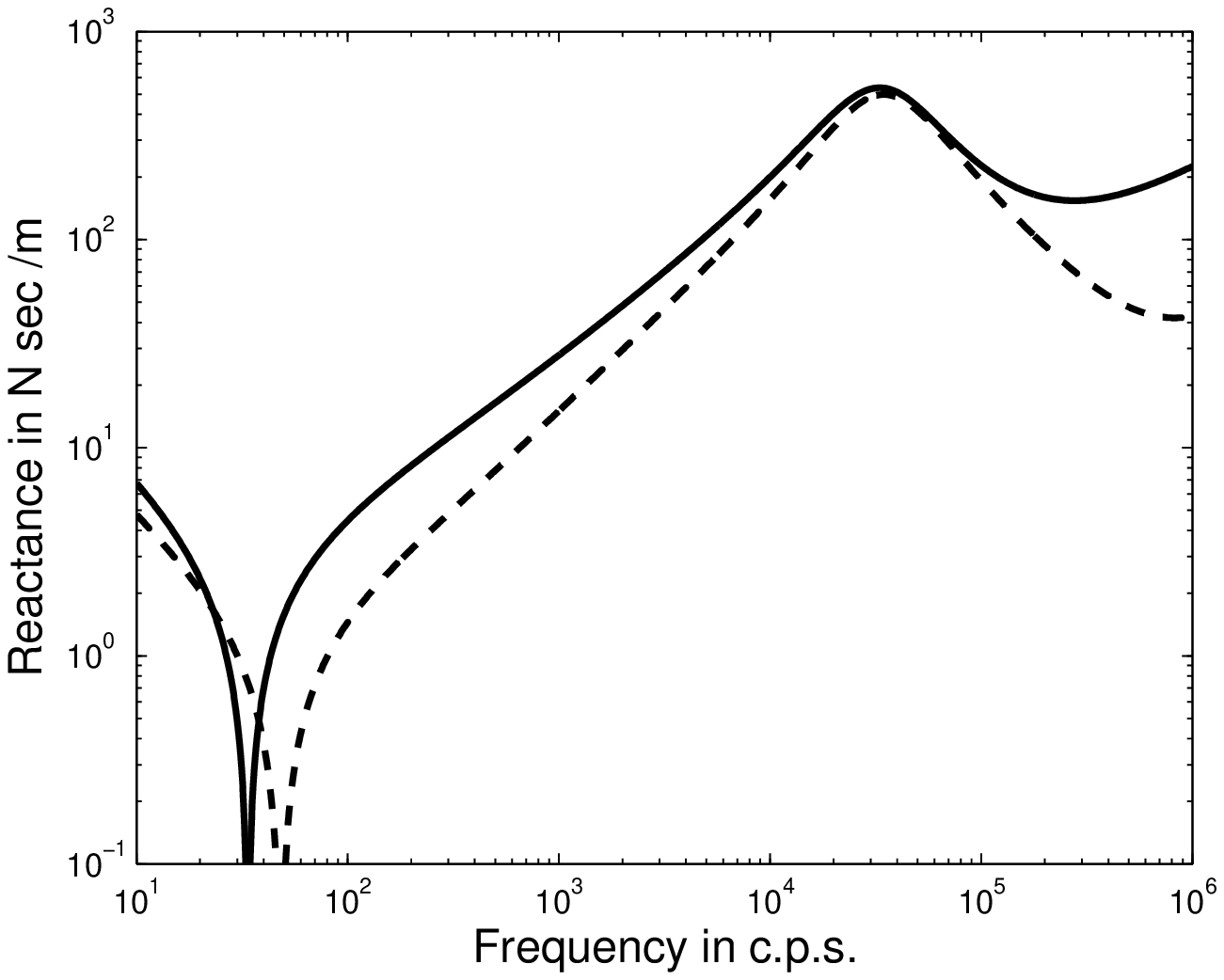}
	\caption{The reactance (real part of $Z$) for an oscillating sphere in a tissue-like material \cite{Oestreicher51}. The solid and dashed curves correspond to a bonded ($\chi = 0$) and  slipping ($\chi = 1$) spherical interface, respectively. The reactance is positive (mass-like) except for frequency below 30 c.p.s. (50  c.p.s. for the dashed curve) where it is negative (stiffness-like).}
		\label{fig1} \end{center}  
	\end{figure}

\begin{figure}[htbp]
			\centering
				\includegraphics [width=5in, height=4 in ] {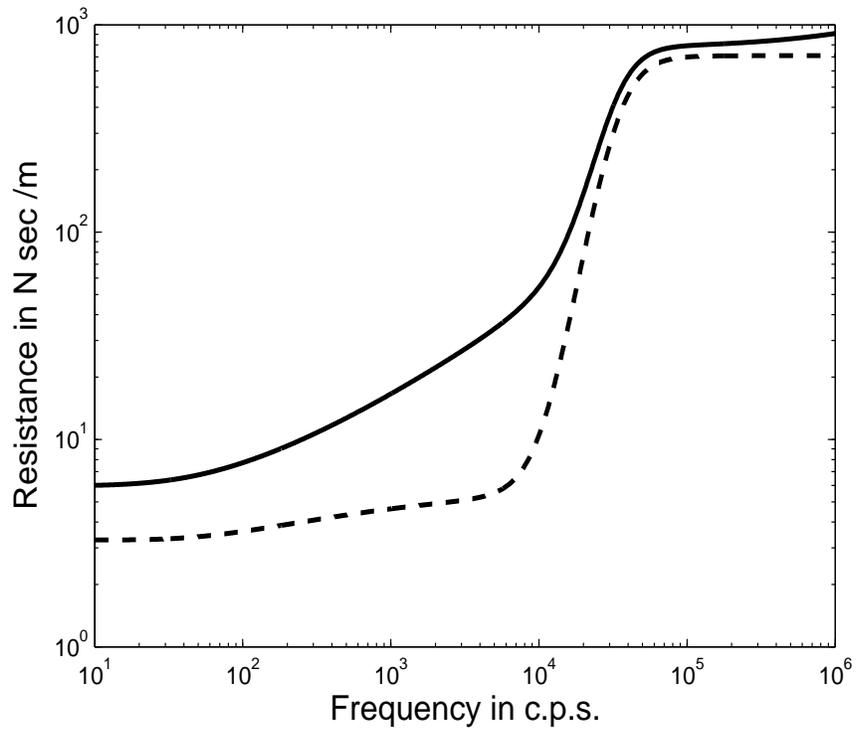}
	\caption{The resistance (imaginary part of $-Z$) for an oscillating sphere in a tissue-like material \cite{Oestreicher51}. The solid and dashed curves correspond to a bonded ($\chi = 0$) and  slipping ($\chi = 1$) spherical interface.  }
		\label{fig2}
	\end{figure}

\end{document}